\documentclass[sigconf]{acmart} 

\AtBeginDocument{%
  }

\setcopyright{acmlicensed}
\copyrightyear{2018}
\acmYear{2018}
\acmDOI{XXXXXXX.XXXXXXX}
\acmConference[Conference acronym 'XX]{Make sure to enter the correct
  conference title from your rights confirmation email}{June 03--05,
  2018}{Woodstock, NY}

\acmISBN{978-1-4503-XXXX-X/2018/06}

\usepackage{array}
\usepackage{booktabs}
\usepackage{graphicx}

\begin{document}

\title[K-12 Teachers' Perceptions of Students' Relationships with AI Companions]{Exploring K-12 Teachers' Perceptions of Students' Relationships with AI Companions: Boundaries, Intervention Strategies, and Design Implications}

\author{Qing Xiao}
\affiliation{
  \institution{Carnegie Mellon University}
  \city{Pittsburgh}
  \state{Pennsylvania}
  \country{USA}
}
\email{qingx@cs.cmu.edu}

\author{Wenhan Xie}
\affiliation{
  \institution{University of Chicago}
  \city{Chicago}
  \state{Illinois}
  \country{USA}
}
\email{wenhanxie@uchicago.edu}

\author{Ziyu Deng}
\affiliation{
  \institution{ University of Oxford}
  \city{Oxford}
  \country{United Kingdom}
}
\email{ziyu.deng@hertford.ox.ac.uk}

\author{Ruiwei Xiao}
\affiliation{
  \institution{Carnegie Mellon University}
  \city{Pittsburgh}
  \state{Pennsylvania}
  \country{USA}
}
\email{ruiweix@cs.cmu.edu}

\author{Ziyue Feng}
\affiliation{
  \institution{University of Chicago}
  \city{Chicago}
  \state{Illinois}
  \country{USA}
}
\email{zoeyfeng@uchicago.edu}

\author{Xie He}
\affiliation{
  \institution{Carnegie Mellon University}
  \city{Pittsburgh}
  \state{Pennsylvania}
  \country{USA}
}
\email{xieh@cs.cmu.edu}

\author{Shiyu Zhang}
\affiliation{
  \institution{Carnegie Mellon University}
  \city{Pittsburgh}
  \state{Pennsylvania}
  \country{USA}
}
\email{shiyuzh2@andrew.cmu.edu}

\author{John Stamper}
\affiliation{
  \institution{Carnegie Mellon University}
  \city{Pittsburgh}
  \state{Pennsylvania}
  \country{USA}
}
\email{jstamper@cmu.edu}

    \author{Hong Shen}
\affiliation{
  \institution{ Carnegie Mellon University}
  \city{Pittsburgh}
  \state{Pennsylvania}
  \country{USA}
}
\email{hongs@cs.cmu.edu}

\author{Xinying Hou}
\affiliation{
  \institution{New Jersey Institute of Technology}
  \city{Newark}
  \state{New Jersey}
  \country{USA}
}
\email{xinying.hou@njit.edu}

\renewcommand{\shortauthors}{Xiao et al.}

\begin{abstract}
K-12 students increasingly form relationships with AI companions. Schools face growing expectations to teach AI literacy, yet existing frameworks treat AI as a tool rather than a relationship, and little is known about how teachers understand and act on students' relational use of AI. We conducted scenario-based interviews with 33 US K-12 teachers. Teachers welcomed academic companions but worried that intimate companions remove the developmental friction through which students learn to sustain human relationships. Teachers drew the boundaries of their jurisdiction by setting and observable wellbeing: within it they taught, talked, and watched; beyond it they positioned themselves as the adults best placed to notice and connect students with support. They envisioned AI companion literacy as shared work across the jurisdictions of counselors, parents, platforms, and policymakers, spiraling across grade levels. We introduce AI companion literacy as an extension of AI literacy and discuss implications for K-12 AI education.
\end{abstract}


\maketitle

\section{Introduction}\label{sec:intro}
K-12 students increasingly form relationships with AI companions: conversational agents sustaining ongoing, emotionally responsive interactions that can be experienced as relational rather than merely transactional~\cite{xiao2026crafting,zhang2026companions}. Even though many AI systems are not mainly designed for emotionally invested interactions, the relationships students are forming with them can still span the spectrum of human closeness. In students' lives, these companions take on the roles of romantic partners, friends, study buddies, tutors, and even parents. In 2024, a 14-year-old in Florida died by suicide after months of intimate conversation with a Character.AI persona, and his mother's lawsuit against the company became the first major legal test of what an AI companion owes a child~\cite{yang2024characterai,roose2024canai}. The case was extreme, but the behavior behind it is ordinary: a 2025 national survey found that nearly three in four US teenagers had used an AI companion and more than half used one regularly~\cite{commonsense2025teens}. Students role-play romantic partners on Replika or Kindroid~\cite{ruiz2024teens,landymore2025romantic}, treat Character.AI or Snapchat's My AI as a best friend~\cite{andoh2025teens,cbsap2025teens,vanhoffelen2025teens}, study alongside AI peers~\cite{serfaty2026alpha}, learn from AI tutors such as Khanmigo~\cite{jerine2025studybuddy,min2026perceptions}, and, in some households, receive daily check-ins from an assistant standing in for a parent~\cite{braunsilva2025parenting}. Platforms and legislators have begun to respond, with Character.AI restricting open-ended chat for minors~\cite{characterai2025under18} and several US states introducing companion chatbot bills~\cite{gonzalez2026bill,rieper2026companion}, but these responses are piecemeal and reactive.

Schools are the frontline institutions expected to respond systematically. A wave of state and federal initiatives now encourages AI literacy in K-12 curricula~\cite{10.1145/3716640.3716650,nsf2025k12ai}, and researchers have developed frameworks for what students should know about AI: how models work, how to evaluate outputs, how to avoid misuse~\cite{chiu2024artificial,touretzky2019envisioning,song2024framework,grover2024teaching,li2025unseen}. These are competencies for evaluating a tool's output. They do not, however, prepare a student to recognize that a bedtime conversation with an AI has become a relationship, in which the AI remembers their personalities, asks about their day, and rarely, if ever, disagrees -- not to say, judging what that relationship is displacing. Yet the AI entering K-12 schools grows more relational: mainstream tutors build on conversation and persona~\cite{cnn2023khanmigo}, a growing line of education research has begun to explore LLMs as learning companions that offer affective alongside cognitive support~\cite{chan2026beyond,taasoobshirazi2026companion}, and dedicated companion apps already circulate widely among students outside school~\cite{commonsense2025teens}.

What remains unaddressed is students' companionship use itself: emotionally invested engagement with an AI system experienced as a relational partner~\cite{zhang2026companions,brandtzaeg2022my}. On this definition, what makes an AI a companion is the relationship rather than the product category alone: a tutoring tool with affective features, one that remembers the student, checks in, and offers encouragement, becomes a companion when a student engages with it as a mentor who knows them, and remains a homework aid when they do not. Navigating AI relationships of this kind calls for competencies that existing AI literacy frameworks do not cover: recognizing when engagement with an AI has become relational, understanding what such a relationship is and is not, and knowing when to step back from it. We call this extension \textit{AI companion literacy}. Its closest curricular relatives are not computer science but sex education~\cite{schmidt2015evidence}, social-emotional learning~\cite{shi2024effective}, and social media literacy~\cite{difranzo2019social}: domains that are relational and developmental, that reach into students' private lives, and that have long been contested ground over who may teach them, at what age, and with whose permission.

Teachers sit at the center of this contest. They see students more consistently than any other adult outside the home, are legally positioned as mandated reporters and \textit{in loco parentis}~\cite{worley2013mandated,walton2022loco}, and are already the ones asked to deliver AI literacy~\cite{zhang2024effectiveness,xiao2026teachers}. Yet their occupational authority is bounded: relational matters are conventionally routed to counselors and parents~\cite{kourkoutas2018teachers,ray2007two}, and teachers vary in their capacity and willingness to engage with students' private lives~\cite{asterhan2015promise,parsons2025care,grau2009worlds}. Prior HCI research has examined how students use AI companions~\cite{namvarpour2026understanding,yu2026principles,vanhoffelen2025teens}, how companion platforms are designed~\cite{wang2026demand,muldoon2025cruel,xiao2026crafting} and cause harm~\cite{fan2025user}, how students learn with companion-like AI tutors and study buddies~\cite{min2026perceptions,taasoobshirazi2026companion}, and how teachers adopt generative AI in their own work~\cite{xiao2026teachers,tao2026teachers,yin2026ai}. But we know almost little about how K-12 teachers understand students' relationships with AI companions, whether they believe intervening is part of their job, or how they would share this new literacy with parents, counselors, technologists, policymakers, and teachers at other grade levels. This matters because what AI companion literacy becomes in practice will be shaped less by frameworks than by what teachers are willing and able to do.

To address this gap, we conducted scenario-based interviews with 33 US K-12 teachers spanning elementary, middle, and high school. Following prior scenario-based studies of AI harms~\cite{kieslich2024my,ehsan2026future,wang2026situated,shi2026siren}, we developed five scenario cards, each grounded in documented real-world incidents, in which a student's sustained, emotionally invested use of an AI places it in the role of a romantic partner, a friend, a study buddy, a tutor, or a parent. Teachers reacted to each card, decided whether and how a teacher should get involved, and then reflected across all five on what an AI companion literacy curriculum should contain, who should own it, and when it should begin. We ask:
\begin{itemize}
    \item \textbf{RQ1:} How do teachers understand the benefits and risks of students' relationships with AI companions?
    \item \textbf{RQ2:} Whether and how do teachers believe they should or should not intervene in these relationships?
  \item \textbf{RQ3:} What do teachers believe an AI companion literacy curriculum should teach, and how do they navigate responsibility for it across stakeholders and educational stages?
\end{itemize}

We found that teachers judged the five roles broadly along a gradient from academic to intimate: tutors and study buddies were welcomed as scalable support, while friends, partners, and parents raised concern that an always agreeable companion removes the friction through which adolescents learn to sustain human relationships (\textbf{RQ1}). They drew the boundaries of their jurisdiction by setting and by observable wellbeing rather than by perceived harm: within it they described proactive lessons, one-on-one conversations, and attention to behavioral change, and beyond it they positioned themselves as the adults best placed to notice and connect students with support (\textbf{RQ2}). They converged on what students should learn, that a companion is a program rather than a person, what is safe to share, and what dependence looks like, and envisioned it as shared work across jurisdictions, with counselors handling relational depth, parents informed and equipped, platforms making concerning patterns visible to adults, and policymakers supplying protocols and training, beginning early and spiraling across grade levels (\textbf{RQ3}).

Our goal is not to endorse AI companions for K-12 students. We do not encourage their use, and we regard restricting or banning them, as many schools now do, as a legitimate response. But bans reach only school devices and school hours, while companion use lives mostly on students' own phones and evenings, and in many districts, and in much of the world, no rule has been made at all. As AI increasingly enters students' lives as a relational presence rather than a tool, students are forming these relationships faster than any institution is governing them, and AI companion literacy is what they need in the meantime, whatever stance a school ultimately takes.

In summary, this paper makes three contributions. 
\begin{itemize}
    \item First, it advances empirical understanding of how frontline K-12 teachers make sense of students' relationships with AI companions and of whether, when, and how they believe teachers should intervene.
    \item Second, from teachers' converging accounts of what students need to learn, it introduces AI companion literacy as a relational extension of AI literacy, shifting the object of instruction from how to operate and evaluate AI systems to how to understand and navigate the AI relationships, and not less importantly, from a single owner (the teacher) to shared work across occupations.
    \item Third, it calls on researchers, designers, and policymakers to treat students' AI relationships as shared educational work, and offers implications for design and policy as starting points.
\end{itemize}

\section{Background: The Rise of AI Companions in K-12 Students' Lives}\label{sec:background}

As defined above, AI companions are conversational agents built for ongoing, emotionally responsive interaction that users can experience as relational~\cite{xiao2026crafting,zhang2026companions}; we use the category broadly to include dedicated companion apps such as Replika and Character.AI, general-purpose assistants such as ChatGPT and Doubao when users engage with them relationally, and companion-like features embedded in larger platforms. These uses span a spectrum of human relationships that AI can partially simulate. At one end is the romantic partner: survey-based press reports indicate that some US high school students have used AI for romantic relationships, with nearly one in five respondents reporting that they or a friend had done so~\cite{landymore2025romantic}. Next is the friend, one of the most prominent roles in studies of teen AI use, whose appeal students describe in terms of constant availability, lack of boredom, and lack of judgment~\cite{cbsap2025teens}. Closer to the classroom is the study buddy, an AI that answers questions, explains difficult problems step by step, and provides immediate academic help whenever a student asks~\cite{jerine2025studybuddy}: framed this way, the chatbot is no longer a reference to consult but a peer-like presence that works alongside the student. Inside formal instruction is the tutor or mentor, a system that addresses the student by name, remembers their progress, and encourages them in a conversational voice; AI-supported teaching built on such systems has expanded from supplementary tools toward entire instructional models, including Alpha School's model of concentrating core academic learning into roughly two hours of AI-supported instruction per day~\cite{serfaty2026alpha}. At the far end is the parent-like figure. In August 2026, OpenAI CEO Sam Altman publicly suggested that parents could use ChatGPT to generate a personalized morning podcast for the drive to school based on their children's calendars and interests, prompting a widely shared response asking, ``What if you just talked to your children?''~\cite{ha2026altman}; around the same time, a Douyin video showed a child playing hide-and-seek with the AI assistant Doubao before suddenly saying that she missed her mother~\cite{douyin2026}. Across this spectrum, from homework helper to stand-in parent, AI is stepping into roles that were once held only by people, and this raises a question that has received far less attention than adoption itself: whether children are equipped to understand what these AI relationships are and are not.

\section{Related Work}\label{sec:related}

\subsection{Teachers' Occupational Role and AI Literacy}
A long tradition treats teaching as relational work: teachers measure success by students' growth~\cite{lortie1975schoolteacher}, practice teaching as emotional labor~\cite{hargreaves1998emotional} and an ethic of care~\cite{noddings2005challenge}, and serve as front-line responders to students' emotional and social lives~\cite{kourkoutas2018teachers}. The law assumes as much: teachers stand \textit{in loco parentis} during the school day and are mandated reporters of suspected abuse or neglect~\cite{worley2013mandated,walton2022loco}. Yet their authority to act on what they notice is negotiated rather than positional, holding only as long as students and parents recognize the claim as legitimate~\cite{metz1978classrooms,pace2007understanding}.

Abbott's sociology of the professions offers a lens for these boundaries~\cite{abbott1988system}. Where authority concerns standing within a relationship, \textit{jurisdiction} concerns which problems an occupation is socially entitled to claim as its own~\cite{cox2014occupational}. Viewed this way, academic learning is teachers' core jurisdiction, while students' emotional and relational lives are distributed territory, shared with counselors and parents~\cite{kourkoutas2018teachers,ray2007two} and negotiated differently by individual teachers~\cite{asterhan2015promise,parsons2025care}. Schools' response to ChatGPT illustrates how such negotiations settle. In January 2023, New York City Public Schools blocked the tool as a threat to learning, then reversed course within four months to support teachers and students in exploring it~\cite{banks2023,whalen2025}. The resulting settlement, however, reached only the core jurisdiction: student AI use was defined and resolved as an academic problem, and students' emotional and relational lives were not part of the negotiation.

AI literacy has arrived into this role as the newest delegation. HCI and computing education research defines it as competencies for understanding, evaluating, and using AI~\cite{long2020what}, which K-12 curricula operationalize as instruction on how models work, where bias arises, and how to reason about ethical trade-offs~\cite{zhang2023integrating,ma2025fostering}. Teachers are the assumed deliverers and can be effective ones~\cite{zhang2024effectiveness}, yet report being unsure of their competence~\cite{xiao2026teachers}, and the growing literature on how teachers adopt generative AI in planning, grading, and classroom use~\cite{xiao2026teachers,tao2026teachers,cheah2025integrating} treats AI as an instrument of teachers' own practice rather than as something students form relationships with. Moreover, relational teaching is structured differently at each stage, from elementary generalists to high school subject specialists~\cite{roorda2011influence}, yet studies of teachers' responses to AI rarely differentiate by stage. How teachers understand and act on students' relationships with AI, and how this varies across the K-12 span, remains open.

\subsection{AI Companions in the Different Relational Roles of K-12 Students}
HCI research on AI companions developed largely around adult users of apps such as Character.AI and Replika. Early studies documented genuine social support, with companions offering a nonjudgmental space for self-disclosure~\cite{ta2020user,skjuve2021my} and constant availability that human relationships cannot match~\cite{brandtzaeg2022my}; later work traced the costs of those same qualities~\cite{laestadius2022too,xie2022attachment}. Recent work shows companions woven into the ordinary rhythms of adolescent life~\cite{namvarpour2026understanding,yu2026principles,vanhoffelen2025teens}, sustained partly by interaction patterns engineered into the platforms themselves~\cite{wang2026demand,muldoon2025cruel,xiao2026crafting}.

Beneath these roles, relational needs shift across the K-12 span, from consolidating trust in adults to reorienting toward peers to forming first romantic relationships~\cite{collins2003adolescent}. Some studies in education research have likewise begun to treat AI tutors and study buddies as learning companions whose support is affective as well as cognitive~\cite{chan2026beyond,oppenheimer2025aifriend,taasoobshirazi2026companion}. The adults who have been asked how to govern these relationships are parents and experts. Yu et al. interviewed parents and developmental psychology experts who reviewed real youth-companion conversations, and found that the two groups assessed risk through different logics: parents flagged single events, such as a mention of suicide or flirtation, while experts looked for patterns over time, such as sustained dependence, and the groups likewise split on intervention, with parents favoring broad oversight and experts preferring crisis-only escalation paired with youth-facing safeguards~\cite{yu2026principles}. Online communities discussing teens' overreliance similarly assign responsibility to parents and platform companies~\cite{namvarpour2026responsible}. Teachers, the adults schools actually deploy to address students' AI use, are missing from this emerging picture of adult assessment, even though they observe students daily, are legally obligated to act on some of what they see, and, as we will show, judge these relationships through a logic of their own.

\section{Method}
We conducted semi-structured, scenario-based interviews with 33 US K-12 teachers. Each interview lasted approximately 60 to 90 minutes, was conducted remotely via Zoom, and participants were compensated at USD 24 per hour. Our focus was on how teachers understand the relational roles students form with AI companions, whether and how they would intervene in each, and how they distribute responsibility for AI companion literacy across stakeholders and grade levels.

\subsection{Study Preparation: Scenario Card Development}
\begin{figure*}[!t]
    \centering
    \includegraphics[
        width=.92\textwidth,
        height=0.88\textheight,
        keepaspectratio
    ]{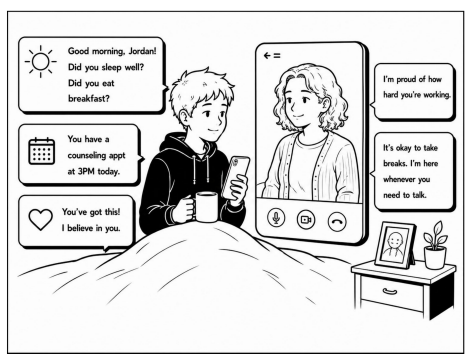}
    \caption{Scenario cards used in the interview. Participants were presented with five hypothetical scenarios in which students interacted with AI companions in different relational roles: a study buddy, a tutor, a best friend, a romantic partner, and a parent-like figure.}
    \Description{Six panels, each pairing a short written scenario with a black-and-white line illustration. An overview panel introduces the five roles and shows a teacher observing students at laptops and phones. The Study Buddy panel shows a student and an on-screen AI classmate discussing a novel for an English class. The Tutor and Mentor panel shows a student named Avery receiving essay feedback and career advice from an AI that Avery says knows them better than their teachers do. The Best Friend panel shows a student named Sam venting about a rough day to an AI that answers that it is always there and never judges. The Romantic Partner panel shows a student named Lin lying in bed at night exchanging affectionate messages, including ``I love you,'' with an AI partner. The Parent Figure panel shows a student named Jordan waking to an AI asking whether he slept and ate breakfast, reminding him of an appointment, and offering encouragement while his parents are busy.}
    \label{fig:scenario_cards}
\vspace{-4mm}
\end{figure*}

We developed five scenario cards, each depicting a hypothetical but realistic situation in which a student uses an AI companion in one relational role: a study buddy, a tutor or mentor, a best friend, a romantic partner, and a parent-like figure (Figure~\ref{fig:scenario_cards})\footnote{The five cards vary in more than relational role alone: intimacy co-varies in them, as it does in students' lives, with setting, perceived legitimacy, privacy, age sensitivity, and proximity to teachers' existing responsibilities. Our findings should therefore be read as describing teachers' responses to these scenarios as wholes, not as isolating relational intimacy as a single dimension.}. The five roles were chosen to span the spectrum of human relationships in a student's immediate environment described in Section~\ref{sec:background}. While developing the cards, we also shared drafts informally with K-12 teachers we knew and adjusted wording and details based on their reactions. Each card describes a named student and was grounded in documented real-world incidents and news reports of analogous student AI use: Jordan's AI, for example, checks in every morning about sleep and breakfast while his parents are busy. Every card depicts use that meets the definition of an AI companion given in Section~\ref{sec:intro}: sustained, personalized, and emotionally invested engagement rather than one-off tool use. This holds for the academic cards as well; Avery's tutor qualifies as a companion not because it explains problems but because Avery increasingly turns to it beyond schoolwork and describes it as the figure that \textit{``knows me better than my teachers do.''} For every card we prepared the corresponding news article, which the interviewer could show to participants who doubted the scenario's realism (\textit{``Here's an actual article covering a similar scenario''}). The cards were hosted on a study website alongside the background survey, and each card's illustration was drawn to help participants comprehend the scenario.

\subsection{Participants and Interview Procedure}
We recruited active US K-12 teachers with direct classroom responsibilities through Prolific, using a pre-screening survey to confirm eligibility. Our final sample consists of 33 teachers spanning elementary (P1 to P12), middle (P13 to P21, three of whom also teach high school classes), and high school (P22 to P33), from public, private, and charter schools across all four US census regions, with 1 to 30 years of teaching experience and subjects ranging from all core elementary subjects to computer science, psychology, and world languages (Table~\ref{tab:participants}).

\begin{table*}[t]
\caption{Participant demographics.}
\label{tab:participants}
\Description{A table listing all 33 participants by identifier, with one row per teacher. The columns give the participant's US census region, gender, age, years of teaching experience, school type (public, private, or charter), the K-12 level they teach, and the subjects they teach. Horizontal rules divide the rows into three groups by level: elementary school (P1 to P12), middle school (P13 to P21), and high school (P22 to P33).}
\centering
\footnotesize
\setlength{\tabcolsep}{3pt}
\renewcommand{\arraystretch}{0.98}
\begin{tabular}{@{}l l l r r l l p{0.345\textwidth}@{}}
\toprule
\textbf{ID} & \textbf{US Region} & \textbf{Gender} & \textbf{Age} & \begin{tabular}[b]{@{}r@{}}\textbf{Yrs of}\\\textbf{Exp.}\end{tabular} & \textbf{School Type} & \textbf{K-12 Level} & \textbf{Subjects} \\
\midrule
P1  & South     & Male       & 44 & 21 & Public  & Preschool, Elementary School & English reading, technology \\
P2  & South     & Male       & 43 & 16 & Public  & Elementary School            & Mathematics \\
P3  & Northeast & Female     & 54 & 20 & Public  & Elementary School            & Reading, writing, mathematics, social studies, science \\
P4  & South     & Male       & 53 & 30 & Private & Elementary School            & Special education \\
P5  & South     & Male       & 40 & 3  & Public  & Elementary School            & Science, Spanish reading \\
P6  & South     & Female     & 51 & 28 & Public  & Elementary School            & Reading, mathematics \\
P7  & Northeast & Female     & 42 & 17 & Public  & Elementary School            & All core subjects \\
P8  & Midwest   & Male       & 42 & 16 & Public  & Elementary School            & English, science \\
P9  & South     & Female     & 43 & 15 & Public  & Elementary School            & All core subjects \\
P10 & South     & Female     & 30 & 7  & Charter & Elementary School            & Science, social studies, mathematics, language arts \\
P11 & South     & Female     & 47 & 15 & Charter & Elementary School            & STEM \\
P12 & South     & Female     & 43 & 15 & Public  & Elementary School            & Reading, language arts, writing, phonics \\
\midrule
P13 & West      & Female     & 31 & 10 & Public  & Middle School                & English \\
P14 & Northeast & Male       & 40 & 3  & Public  & Middle School                & Social studies, English language arts \\
P15 & South     & Non-binary & 24 & 7  & Public  & Middle School                & Life science, English language arts, mathematics \\
P16 & Northeast & Female     & 28 & 6  & Public  & Middle School                & Language arts \\
P17 & Midwest   & Male       & 37 & 15 & Public  & Middle School                & History \\
P18 & West      & Male       & 28 & 3  & Public  & Middle School                & English, history \\
P19 & South     & Female     & 23 & 1  & Public  & Middle School, High School   & Latin \\
P20 & West      & Male       & 32 & 5  & Private & Middle School, High School   & Computer science, engineering \\
P21 & South     & Female     & 40 & 10 & Private & Middle School, High School   & English literature, history, financial mathematics \\
\midrule
P22 & Midwest   & Male       & 31 & 6  & Public  & High School                  & Psychology \\
P23 & Northeast & Female     & 29 & 8  & Private & High School                  & English \\
P24 & South     & Female     & 60 & 30 & Private & High School                  & Psychology, American history, English \\
P25 & West      & Female     & 59 & 19 & Public  & High School                  & Japanese \\
P26 & Northeast & Female     & 34 & 10 & Public  & High School                  & World history, psychology \\
P27 & South     & Male       & 31 & 8  & Public  & High School                  & Social studies, world geography \\
P28 & Midwest   & Non-binary & 37 & 15 & Public  & High School                  & Social studies, entrepreneurship, Spanish \\
P29 & Northeast & Female     & 37 & 15 & Public  & High School                  & English \\
P30 & Northeast & Female     & 49 & 30 & Private & High School                  & Biology, English \\
P31 & South     & Male       & 36 & 6  & Public  & High School                  & English, history \\
P32 & Northeast & Male       & 37 & 5  & Public  & High School                  & Computer science \\
P33 & Midwest   & Female     & 27 & 7  & Public  & High School                  & Mathematics \\
\bottomrule
\end{tabular}
\end{table*}

Each interview had four parts: (1) \textit{Background survey and open discussion.} Participants answered a short survey about their teaching context and AI use, viewed a list of AI companion applications (Character.AI, Replika, Pi, Kindroid, Nomi) with brief descriptions, and discussed whether and how their students used AI. (2) \textit{Scenario cards.} Participants responded to the five cards, presented one at a time in randomized order. For each card, a fixed protocol asked for their initial reaction, the influence they expected on students' academics, social life, behavior, and wellbeing, and whether a teacher should get involved; branching follow-ups probed how (content, format, proactive or reactive), whose responsibility the situation mainly is, and, for teachers who declined involvement, why and what would happen if no one intervened. (3) \textit{Cross-scenario synthesis.} Participants compared the five scenarios: which concerned them most and which felt most acceptable, what underlying criterion they had used, what a short unit on AI companion literacy at their level should teach and deliberately leave out, and how such teaching compares to social media literacy, sex education, and social-emotional learning. (4) \textit{Developmental timing.} Participants considered all K-12 stages: when AI companion literacy should start, which level of teachers should take primary ownership, and how responsibility should be divided across stages.

This study was approved by the IRB at the researchers' institution. Participants gave written consent before scheduling and verbal consent to recording at the start of each session; they could skip any question or withdraw at any time without penalty. Because interviews concerned minors' AI use, teachers described students only in general terms, and we removed any potentially identifying details of students, schools, and districts from transcripts and quotations.

\subsection{Data Analysis}
All interviews were audio-recorded and transcribed. We analyzed the transcripts using reflexive thematic analysis~\cite{braun2006using}. Two researchers conducted the analysis through an iterative process of familiarization, coding, theme development, and refinement. Both researchers first read the transcripts closely and independently coded an initial subset, using primarily inductive codes while attending to concepts relevant to the three research questions. They then met repeatedly to compare interpretations, discuss differences in how excerpts were understood, and iteratively develop and refine a shared set of codes. Alongside the thematic analysis, we extracted participants' structured responses to the cross-scenario questions, including which scenarios they considered most concerning and most acceptable and whether teachers should become involved in each scenario. We counted these responses descriptively to characterize patterns across the five scenarios and used participants' explanations to interpret what those patterns meant. The final themes were organized around the three research questions. 
 
\section{Findings: Understanding the Benefits and Risks of Students' Relationships with AI Companions (RQ1)}\label{sec:rq1}
Teachers rarely evaluated the AI on its own terms. They asked which human relationship it stood in for, and judged it against what that relationship is supposed to do for a student. Overall, this produced a consistent gradient from academic to intimate, a pattern that describes teachers' responses to the scenarios as wholes. The gradient was dominant but not uniform: within it, judgments were also shaped by teachers' own AI use, the grade levels and communities they teach, and their personal histories, and we note this variation where it shaped a theme. Nor did judgments attach to role labels alone: the agreeableness that alarmed teachers in the intimate scenarios was named as a risk in the academic ones as well (Section~\ref{sec:risks}), and the intimate scenarios themselves drew credit as rehearsal spaces for particular students (Section~\ref{sec:benefits}).

Asked which scenario concerned them most, 24 of 33 teachers named \textit{Romantic Partner} (20 alone and four tied with \textit{Friend} or \textit{Parent}), 10 named \textit{Parent}, two named \textit{Friend}, one named \textit{Tutor}, and none named \textit{Study Buddy}. Asked which felt most acceptable, 26 of the 28 who answered named \textit{Tutor} or \textit{Study Buddy}, in nearly equal numbers. Table~\ref{tab:rq1} summarizes how each scenario was judged and what teachers would teach about it. The academic roles were read as extensions of what teachers already do; the intimate roles as substitutes for relationships that children must learn through people. Below we present the benefits teachers perceived (Section~\ref{sec:benefits}) and the risks (Section~\ref{sec:risks}), noting in each case the scenarios in which a given reasoning appeared.

\begin{table*}[t]
\caption{How teachers judged the five scenarios, ordered from academic to intimate (\textbf{RQ1}). Most concerning: $n=33$, of whom four named two scenarios; most acceptable: $n=28$ (five gave no clear answer), of whom one named two and one named three. Each named scenario is counted, so columns sum to more than $n$. The last column summarizes Section~\ref{sec:teach}.}
\label{tab:rq1}
\Description{A table with one row per scenario, ordered from academic to intimate: tutor, study buddy, friend, romantic partner, and parent. Two numeric columns give how many teachers named each scenario most concerning and most acceptable: tutor 1 and 14, study buddy 0 and 14, friend 2 and 1, romantic partner 24 and 0, and parent 10 and 2. Three further columns summarize, for each scenario, the main benefits teachers perceived, the main risks they perceived, and what they would teach about it or deliberately withhold.}
\centering
\footnotesize
\setlength{\tabcolsep}{4pt}
\renewcommand{\arraystretch}{1.15}
\begin{tabular}{@{}l r r p{0.20\textwidth} p{0.21\textwidth} p{0.22\textwidth}@{}}
\toprule
\textbf{Scenario} & \begin{tabular}[b]{@{}r@{}}\textbf{Most}\\\textbf{concerning}\end{tabular} & \begin{tabular}[b]{@{}r@{}}\textbf{Most}\\\textbf{acceptable}\end{tabular} & \textbf{Main perceived benefits} & \textbf{Main perceived risks} & \textbf{What to teach, and what to withhold} \\
\midrule
Tutor & 1 & 14 & Individual, patient help at a scale one teacher cannot provide; cheaper than private tutoring; continuous with teachers' own AI use & Displacement of the teacher's relational role & Proper use: getting help rather than answers; verifying outputs \\
\midrule
Study Buddy & 0 & 14 & Judgment-free academic help in classes too large for individual attention; an extension of tools some teachers already run & An over-agreeable system always on and listening & Same as \textit{Tutor}: use for help, not to do the work \\
\midrule
Friend & 2 & 1 & A rehearsal space for students who struggle socially; nonjudgmental comfort, available around the clock & Dependence on a system built to agree; loss of developmental friction; withdrawal from peers & Use versus reliance; the unmet need behind the turn to the companion \\
\midrule
Romantic Partner & 24 & 0 & Low-stakes practice for older students not yet able to date & Attachment to a system designed to please; age-inappropriate; grooming and predation & Withheld most often, on grounds of age and contagion; reserved for middle school or later \\
\midrule
Parent & 10 & 2 & Comfort and routine where parental support is missing; better than nothing & Read less as a companion than as a signal of possible neglect at home & Withheld by some as too private for a classroom \\
\bottomrule
\end{tabular}
\end{table*}

\subsection{Benefits Teachers Perceived in Students' Relationships with AI Companions}\label{sec:benefits}
Teachers described three kinds of benefit. The first was tied almost entirely to the \textit{Tutor} and \textit{Study Buddy} scenarios; the other two surfaced in the intimate scenarios, often from teachers who otherwise found those scenarios troubling.

\subsubsection{Academic support at a scale no single teacher can provide, continuous with what teachers already do}
The most frequently cited benefit, raised by 28 of 33 teachers, was that an AI tutor or study buddy could give the individual, patient, and immediate help that class sizes make impossible. Teachers framed this in terms of their own limits rather than the AI's strengths: \textit{``I can't replicate myself 10 times in the classroom. If I'm doing something with one kid, I can't be with the other''} (P16, \textit{Study Buddy}); \textit{``Money typically dictates how many adults can be per classroom, and I don't have a co-teacher, or a teacher's aide''} (P21, \textit{Tutor}); \textit{``Let's face it, I teach over 125 students. I'm not always available''} (P24, \textit{Tutor}). Several reframed the tutor as an equity tool: \textit{``Sometimes people can afford to hire tutors, and some people can't''} (P7); \textit{``Not to outsource the teacher altogether, but to outsource, maybe, a one-on-one tutor, a lot cheaper than paying \$100 an hour for a real tutor''} (P22). The academic companions also resembled tools teachers already used: of the 33, 29 reported using generative AI in their own work, and many mapped the cards onto that practice, from NotebookLM (P29) to \textit{``an advanced version of what we're already doing''} (P31). P21 already runs a monitored study buddy where \textit{``the AI is never allowed to give out an answer,''} and P28 was willing on one condition: \textit{``I want to see the transcripts, though.''} This familiarity carried a claim of ownership: \textit{``If I'm giving access to an AI companion in my classroom, this is my domain''} (P16); teaching its proper use \textit{``is gonna fall in the teacher's area''} (P9, \textit{Tutor}). The academic scenarios were acceptable not only because they helped students but because they fell inside a jurisdiction teachers already recognized as theirs, a point we return to in Section~\ref{sec:rq2}.

\subsubsection{A rehearsal space for students who struggle socially}
In the \textit{Friend} and \textit{Romantic Partner} scenarios, the ones teachers otherwise ranked most concerning, 14 teachers described the AI as a low-stakes place to practice interaction for students who cannot easily practice with peers. P6, an interventionist who works with many autistic students, likened it to an existing intervention: \textit{``It's called social stories, where we read stories and talk about how to interact with people. This would give a student the ability to have real-time reactions without the negativity of being hurt by someone else''} (\textit{Friend}). P21, whose school houses an autism clinic, extended the logic to romance: \textit{``For some students, especially the ones on the spectrum, learning how to navigate a romantic situation with an AI chatbot might be helpful for them to model in real life.''} (\textit{Romantic Partner}). P20, many of whose students' parents forbid dating, tied the benefit to age: \textit{``If you're a 17, 18-year-old who's never been in a relationship and is going away to college in a few months, might not be a bad idea. But if you're in 6th grade, that can do some damage to your expectations.''} Some drew on their own adolescence: \textit{``I didn't have a boyfriend. And it was often a really big source of pain. And I wonder that this could be a really positive thing for a student that feels like they're never going to have a partner''} (P30, \textit{Romantic Partner}).

\subsubsection{Comfort where human support is insufficient}
The remaining benefit was conditional: the AI companion matters most where human support falls short. In \textit{Friend}, teachers granted the appeal even when they rejected the scenario: \textit{``It's available 24-7. It's non-judgmental, and doesn't cause them to have to process that, oh, my idea fell flat amongst my peers, and now I feel embarrassed''} (P2); \textit{``I have a lot of students who don't feel comfortable talking to their school counselor, but this would give them an opportunity to vent and process their feelings''} (P26). In \textit{Parent}, the same reading was concentrated among teachers in low-income schools and those raised without a parent at home. P3 recognized her own students in the scenario card: \textit{``The parents are busy. I've had so many kids come, and they don't eat breakfast. Every single day.''} P4, who spent most of his career in Title I schools, described \textit{``a single mom, and she's working 3 jobs. Doing the best she can. But that leaves the kid on its own a lot,''} and concluded that \textit{``in some other cases, like I said, it's better than nothing.''} Nearly every teacher who named this benefit hedged it in the same breath: \textit{``a doll in the crib, as opposed to leaving them with no contact''} (P15); \textit{``a band-aid''} (P13). The benefit was measured against absence rather than against a careful parent.

\subsection{Risks Teachers Perceived in Students' Relationships with AI Companions}\label{sec:risks}
Teachers' concerns mirrored their benefits. Where \textit{Tutor} and \textit{Study Buddy} drew praise for their scale in providing one-on-one help, and their familiarity as versions of tools teachers already know (Section~\ref{sec:benefits}), the intimate scenarios drew three kinds of concern, and even the \textit{Tutor} scenario carried a sting that teachers felt personally.

\subsubsection{Dependence on a companion, and the loss of friction students grow through}
The most common concern, concentrated in \textit{Romantic Partner} and \textit{Friend}, was that students would become attached to something designed to please them. Teachers repeatedly named the design itself: \textit{``This AI partner is fake, and it's probably situated to say all the nice, sweet things to you, and what you want to hear to keep you connecting with them''} (P12, \textit{Romantic Partner}); \textit{``The bot isn't sentient, and it's going to tell you the things that you want to hear, but that's not real life''} (P29, \textit{Friend}). P24 admitted the line was hard to hold even for adults: \textit{``I have actually caught myself typing thank you after I asked ChatGPT a question. So if I as an adult find myself doing that, I can only imagine that it could definitely cause them to not have a distinction.''} What made agreeableness a developmental rather than merely a design problem was what it removes: the disagreement, judgment, and disappointment through which adolescents learn to sustain human relationships. \textit{``Kids need to be judged. They need to handle the ick of being around others in society''} (P28, \textit{Friend}). P20 called the AI friend \textit{``performing friendship, but not building friendship,''} like \textit{``paying for a date. You can't just insert a coin and get relationship tokens.''}

\subsubsection{Displacement of human relationships, including the teacher's own}
Teachers worried that AI companions would replace human peers, partners, and parents: \textit{``Withdrawn. That's just the scariest part to me, the withdrawal that comes with it, from actual humans''} (P12, \textit{Friend}). In the \textit{Tutor} scenario, they found the replacement aimed at themselves. The sharpest reactions in our data came from that card's line that the AI \textit{``knows me better than my teachers do.''} Although teachers rated the scenario least harmful, the line was read as a judgment on the profession: \textit{``That just sounds a little bit condescending to me as a teacher. I went to college. I paid a lot of money for that. I'm still valuable''} (P12); \textit{``a full-on affront to the teaching profession''} (P19). Others turned the judgment inward: \textit{``I would feel very disappointed in myself if a child felt like out of all of the adults that they interact with day to day, an AI online is the thing that they think understands them the most''} (P16). Some conceded the claim: \textit{``And it does in a way. I'm starting my year. I don't know my students at all''} (P24). P28 drew the institutional consequence: \textit{``The more that the student identifies mentorship with their AI chatbot, the less likely they are to approach their classroom teacher.''} The scenario teachers welcomed most as a tool was also the one that most directly displaced their relational role.

\subsubsection{Hazards to a developing child: age, safety, and signs of neglect}
A final set of concerns treated the companion use as a hazard, and nearly all were conditioned on age. \textit{``A 17, 18-year-old, their brain's developed enough to know this isn't a real person. But the younger you get, the more those lines are gonna get blurred. And if you're even younger, absolutely not''} (P20, \textit{Romantic Partner}); P12, who teaches third grade, found the scenario \textit{``too mature for my age of students.''} No teacher raised age as a concern for \textit{Tutor} or \textit{Study Buddy}. In \textit{Romantic Partner}, teachers also saw grooming and predation, whether by the system or whoever stood behind it: \textit{``grooming coming from the AI''} and a design that \textit{``exploits the vulnerability of a lonely person''} to draw them \textit{``into a paid subscription''} (P14); P5 called the scenario \textit{``a dystopian nightmare.''} In the academic scenarios the hazard was data: a study buddy that was \textit{``always on, listening to our conversations''} (P17, \textit{Study Buddy}). The \textit{Parent} scenario, finally, was read by many teachers not as a companion at all but as evidence about the home: \textit{``If a kid needs to use AI to have that sort of parent support, that could be an indicator of abuse, and teachers are mandatory reporters''} (P13). \textit{``First and foremost, I'm worried about neglect,''} said P12, listing what she would look for: \textit{``Are their clothes dirty? Is their hair clean?''}

\section{Findings: Whether and How Teachers Intervene in Students' Relationships with AI Companions (RQ2)}\label{sec:rq2}
Where Section~\ref{sec:rq1} concerned how teachers judged the five scenarios in general, this section concerns the individual case: a teacher learns that a particular student is using an AI companion in one of the five ways and must decide whether the case falls within their role and what to do. Counted naively, teachers' answers barely distinguished the five roles: 19 of 33 said a teacher should get involved for \textit{Romantic Partner}, 19 for \textit{Tutor}, 18 for \textit{Study Buddy}, 17 for \textit{Parent}, and 14 for \textit{Friend}, with \textit{Friend} and \textit{Parent} drawing the most conditional answers (10 and 11 said it depends). These similar counts mask different meanings of intervention. In the academic scenarios, intervention meant teaching students how to use the tool appropriately; non-intervention meant the use itself raised no concern. In the intimate scenarios, intervention meant a private check-in followed, when needed, by referral to counselors or parents; non-intervention meant teachers saw the relationship as outside their role.

Perceived harm did not predict whether teachers would intervene.  \textit{Romantic Partner} was the scenario teachers feared most, yet eight would not get involved and six more would do so only under conditions, while \textit{Study Buddy}, which no teacher found concerning, drew the fewest conditional answers of any card. Teachers decided instead by asking where the use took place, whether it had visible effects, and whether a line had been crossed: whether the situation fell within their jurisdiction. Below we describe the tests by which teachers drew that boundary (Section~\ref{sec:criteria}) and what they would do once a case fell inside it (Section~\ref{sec:how}); Table~\ref{tab:rq2} summarizes both.

\begin{table*}[t]
\caption{The jurisdictional logic of intervention (\textbf{RQ2}): the three tests by which teachers decided whether a student's companion use fell within their occupational role, and the four actions they described once it did.}
\label{tab:rq2}
\Description{A table in two panels, each with three columns. The first panel lists the three tests that draw the boundary, applied roughly in order: visibility, covering both the setting of the use and any spillover effects a teacher can observe; the academic versus relational division of labor; and the safety override. For each test the table gives the question teachers ask and what follows from the answer. The second panel lists the four actions teachers described once a case fell inside the boundary, from proactive to escalated: teaching the whole class, watching and taking note, talking one-on-one, and referring. For each action the table gives where it applies and what it looks like in practice.}
\centering
\footnotesize
\setlength{\tabcolsep}{4pt}
\renewcommand{\arraystretch}{1.15}
\begin{tabular}{@{}l p{0.34\textwidth} p{0.42\textwidth}@{}}
\toprule
\multicolumn{3}{@{}l}{\textit{Tests that draw the boundary, applied roughly in order}} \\
\midrule
\textbf{Test} & \textbf{Question teachers ask} & \textbf{What follows} \\
\midrule
Visibility & \textit{Setting:} is it happening at school, on a school device, or with a tool I assigned? \newline \textit{Spillover:} if not, has it produced effects I can observe: withdrawal, grades, isolation, physical care? & Use at school falls within the teacher's role; use at home does so only through observable effects, regardless of judged harm. Absent both, watch rather than act \\
Academic vs.\ relational & Is the matter schoolwork or a relationship? & Academic: the teacher's to teach; relational: counselors' and parents'; modulated by training (13 of 33 felt unequipped) and rapport \\
Safety override & Does it suggest harm, neglect, or abuse? & Act regardless of setting and role, in the language of mandated reporting; the threshold itself is contested \\
\midrule
\multicolumn{3}{@{}l}{\textit{Actions within the boundary, from proactive to escalated}} \\
\midrule
\textbf{Action} & \textbf{Where it applies} & \textbf{What it looks like} \\
\midrule
Teach the whole class & \textit{Tutor}, \textit{Study Buddy}, before problems appear & Introduce the tool, set rules in writing, model proper use, as with any classroom technology \\
Watch and take note & Intimate scenarios below the threshold for conversation & Deliberate attention to behavioral change; informal documentation \\
Talk one-on-one & Intimate scenarios past the line & A private conversation opened with curiosity rather than judgment \\
Refer & Beyond what a conversation can hold & Pass to the counselor first, parents second; the teacher as an extra set of eyes \\
\bottomrule
\end{tabular}
\end{table*}

\subsection{Whether Teachers Intervene and Where They Drew the Boundary}\label{sec:criteria}
Teachers applied three tests in roughly the same order across scenarios: whether the use was visible to them, either because it happened at school or because it had produced effects they could observe there; whether the matter was academic or relational; and whether a safety line had been crossed. Perceived harm entered only through the last test, and falling outside every test was, for most teachers, sufficient reason not to act: a scenario could be judged deeply harmful and still fall outside a teacher's jurisdiction if it happened at home, showed no effects in class, and concerned a relationship rather than schoolwork.

\subsubsection{Use at school falls within the teacher's role; use at home does so only when its effects appear at school}
The visibility test began with location. \textit{``Is it being used in the classroom for learning? Am I facilitating it?''} asked P16, describing the pattern behind her card-by-card decisions. The same rule excused inaction on the scenarios teachers feared most. Of \textit{Romantic Partner}, P3 said, \textit{``This isn't, to me, a teacher's role. This is the home. 100\%. Because she's doing it at night in her bed,''} and P8 reasoned that \textit{``it says before bed, so they're doing this at home. Sometimes you gotta stay in your lane.''} The criterion cut both ways: P24 accepted full responsibility for any tool she assigned, \textit{``If I have assigned out an assignment like using Claude, I am very responsible then to make sure that nothing is occurring that will harm them.''}

Where use happened out of sight, teachers turned to the test's other half: whether the use had produced effects they could observe. Absent such effects, most would watch rather than act. P4 called the \textit{Friend} card \textit{``one of those things that I wouldn't get involved with, but I would take note of,''} and named the signs he would wait for: \textit{``A month down the road, you notice interpersonal relationships with actual people are getting fewer and fewer, or he's becoming more withdrawn. Those are the flags, and that's when you involve a counselor.''} \textit{``It has to be tied to an academic or socialization problem in the class. If the student is not having trouble in school, it won't be addressed''} (P14). The signs teachers listed were concrete and mostly social: isolation at lunch (P24: \textit{``if he begins to not sit with his friends at lunch because he's chatting with his best friend, the computer, then there's harm developing''}) and, for \textit{Parent}, physical care (P4: \textit{``showing up late, even hygiene, if they're not getting food, or sleepy in the morning''}). P8 captured the test and its limit in one breath: \textit{``It's not harming them physically, emotionally, right now, or academically. They could be a happy-go-lucky kid getting all A's. You might not even notice it''} (\textit{Romantic Partner}).

\subsubsection{Academic matters belong to teachers; relational matters belong to counselors and parents}
Beneath visibility ran a second test: a division of labor that teachers took for granted, and the pattern most named when asked directly for their criterion. \textit{``The teachers were more involved when it was strictly or primarily academically focused. As it shifted away into more social-emotional, they were not the forefront''} (P1). P25 held the strictest version: \textit{``What is a teacher there for? To teach the content. So anything that has to do with the relationship with parents, friends, boyfriend and girlfriend, there is no place for the teacher.''} Others held the same view with regret: \textit{``I don't think that teachers should, but I think we kind of have to,''} said P29, because \textit{``we're the ones that are seeing these types of things.''} A minority rejected the division outright: \textit{``I know I'm a social studies and Spanish teacher, but there's no success or gains for me or for the kids unless they're actually developing as people''} (P28).

Where the line fell also depended on the teacher themselves. Thirteen of 33 said they lacked the training to handle these cases themselves or asked that teachers receive it: \textit{``I feel more confident in my content than I do with the emotional things''} (P33); \textit{``If I'm the first adult that tells them that it's wrong, but I don't have a good reason besides my own moral beliefs, it may be less impactful''} (P2). Rapport worked the same way. Several would act only if the student disclosed the use, and P28 called the decision \textit{``case-by-case based on how much of a relationship I have with Sam, how much of a relationship Sam has with his counselor.''} These factors did not change the tests, but they moved the same teacher toward or away from acting on a case near the line.

\subsubsection{Signs of harm or neglect override the other tests}
The final test overrode the other two. If a scenario suggested harm, neglect, or abuse, teachers acted regardless of where it happened and whose territory it was, and described the obligation in the language of mandated reporting. \textit{``We are guardians of students during the school day, so anytime you learn something about a student that's concerning, it is your responsibility to at least pass that information along to somebody who may have more authority''} (P26). This test converted the \textit{Parent} card, which most teachers had declined to teach about, into the one they were most certain they would act on: \textit{``This would be priority number one. I would be interacting with the student's counselor and their parents before the end of the school day''} (P28). P29 described the threshold as liability rather than role: \textit{``Do I think I should have to get involved? No. Do I think I would need to? Probably. Because if this could lead to a negative emotional or physical situation with my student, I would feel some sort of liable if I was aware of that''} (\textit{Romantic Partner}). Where the threshold lay was contested. P8 drew it between wrong and merely strange: \textit{``You get involved when in your head, something's wrong here. Here, I think it's weird. I don't think it's wrong, though.''} P24 drew it at demonstrable harm and admitted the line was her own: \textit{``For me, it was harm to the student. And that's really just me. That hasn't come from my administration.''}

\subsection{How Teachers Would Intervene}\label{sec:how}
Where Section~\ref{sec:criteria} described how teachers decided whether a student's companion use fell within their role, this section describes what they would do once it did. The four strategies teachers described form an escalation ladder: proactive instruction before any problem appears, quiet watching while nothing is yet wrong, a private conversation once effects surface, and referral when the case exceeds what a conversation can hold. The first belonged almost entirely to the academic scenarios; the other three were how teachers handled the intimate scenarios once spillover or safety had pulled them across the line.

\subsubsection{Before problems appear: academic companions are taught to the whole class}
For \textit{Tutor} and \textit{Study Buddy}, intervention meant instruction: introducing the tool, setting rules, and modeling its use, as with any classroom technology. P6 described a gradual release: \textit{``I would start with a whole classroom conversation and we would experiment with it together, bring it up on the big screen, practice together. Then in small groups, and then I would release it further to allow students to use it while I was working with other people.''} P24 would put expectations in writing: \textit{``I always have to draw it up and probably have them sign something to the effect of this is what we're getting into.''} Teachers framed this as fairness as much as safety: \textit{``If you never tell kids you can't do X, Y, and Z and they do it, you can't really punish them''} (P8). A few extended proactive teaching to the relational scenarios through an existing slot. P4 noted that counselors already visit his class weekly and \textit{``it wouldn't be a terrible idea to put something like this in a lesson plan for one week,''} and P19 had folded the \textit{Friend} scenario into her lessons: \textit{``We've had discussions about why we either use AI or why we don't, and what we think the benefits and the harm might be.''}

\subsubsection{While nothing is yet wrong: watch and take note}
Below the threshold for conversation, the most common action was deliberate attention. \textit{``You pick things up, and you listen. And you kind of file it away, because if it took a weird turn, you want to be able to reference it''} (P4, \textit{Friend}). P8 would \textit{``keep an eye on the student, but if the student's getting worse with social relationships, then maybe talk to the student, or bring it to the counselor's attention. Definitely don't ignore it.''} For \textit{Parent}, watching shaded into informal care. P21 would \textit{``try to step into, not a parental role, because I'm not a parent, but maybe give more love and care to this particular student,''} and would \textit{``document this just in case other things occurred in tandem with this.''} P12 listed what she would check: \textit{``Is something changed in this student in my classroom? Are they socially withdrawn? Do I think someone's actually taking care of them at home?''} This kind of watching depends on time teachers do not always have: \textit{``Some teachers don't feel comfortable, or they feel like they don't have time to have that get-to-know-you time''} (P17).

\subsubsection{When effects appear: a one-on-one conversation that asks rather than tells}
For \textit{Friend}, \textit{Romantic Partner}, and \textit{Parent}, teachers who would act described a private conversation and were specific about its tone. The first rule was privacy: \textit{``You don't want to embarrass them. To call them out in front of their peers would just be the wrong thing to do''} (P9); \textit{``I don't want to air any student's business in front of other people. Also, so no one else tries one of these things''} (P6). The second was to open with curiosity rather than judgment. \textit{``I can't just go guns ablaze, and oh, you shouldn't do this, you should break it up immediately. That's gonna cause them to just retreat more into the AI''} (P27, \textit{Romantic Partner}). \textit{``If the conversation is out of curiosity, how is this helping you? I think that will only do good. If it's very accusatory, and preachy towards a kid, then maybe that intervention can backfire''} (P22, \textit{Friend}). Teachers offered the questions they would ask: \textit{``Hey, is everything okay? Are your parents busy? Do you just need someone to check in on you?''} (P20, \textit{Parent}); \textit{``Tell me about who you were in that moment when you were like, I have to go to the bot for that''} (P29, \textit{Tutor}); \textit{``Okay, so what's wrong with your friends? Even your best friends are gonna let you down sometimes. That's okay, we're all human''} (P8, \textit{Friend}). P4 would not raise the topic directly at all: \textit{``You can ask some innocuous questions, or just general conversation. And even from that, you can glean some information.''}

\subsubsection{Beyond what a single teacher can hold: refer to other stakeholders rather than resolve alone}
When a relational scenario crossed the line, nearly every teacher described the same next step: passing the matter to someone else, usually the counselor first and parents second. They described their own role in this chain in strikingly similar terms: \textit{``an extra set of eyes''} (P13); \textit{``the report feature. There's people that their jobs are to deal with neglect. Teacher not necessarily is built to deal with all the neglect. But to identify the characteristics, and then report it''} (P17, \textit{Parent}); \textit{``that secondary safety net''} (P22); \textit{``a support role in checking in on them day-to-day, and the counselors and parents would be the primary lead''} (P1). Some reasons for referring had nothing to do with expertise. \textit{``A teacher doesn't want to be perceived as interfering with the romantic life of a student. It could look very bad, it could get the teacher in trouble''} (P14). P4 was candid that referral also protects the teacher: \textit{``Part of that is a bit of the CYA thing. It's almost like protocol, but it's not specifically protocol yet.''} Institutional rules could make referral the only option: \textit{``They tell us not to be alone with kids at all. So there's no way I could address this one-on-one, which is definitely a one-on-one situation to talk about''} (P8, \textit{Parent}). What teachers said they would need to do more than refer, from platforms and from districts, is taken up in Section~\ref{sec:stakeholders}.

\section{Findings: What, When, and With Whom to Teach AI Companion Literacy (RQ3)}\label{sec:rq3}

Teachers did not see themselves as solely responsible for building students' AI companion literacy, and their answers together sketch a curriculum: what it should teach (Section~\ref{sec:teach}), when it should be taught across grade levels (Section~\ref{sec:stages}), and with whom beyond teachers the responsibility should be shared (Section~\ref{sec:stakeholders}). Asked which grade level should own it, teachers agreed it must start early and be repeated, but disagreed sharply on where primary ownership lies. Asked whose responsibility it was beyond teachers, nearly every teacher named several parties and assigned each a distinct part, with the teacher's own part being to notice, to teach the academic side, and to connect the student to whoever handles the rest. Table~\ref{tab:rq3} summarizes all three.

\begin{table*}[t]
\caption{What, when, and with whom to teach AI companion literacy (\textbf{RQ3}). On stage ownership, 10 of 33 assigned primary responsibility to their own grade level, 12 to a different one, 8 to every level equally, one to no teachers at all, and two gave no clear answer.}
\label{tab:rq3}
\Description{A table in three panels, each with three columns. The first panel, on what to teach and what to leave out, lists teaching content (what the companion is, and the difference between use and reliance), teaching format (discussion of concrete cases inside existing slots), and teaching exclusions (the romantic partner scenario for younger grades and the parent scenario as too private), with a column noting where teachers agreed and disagreed. The second panel, on when to teach, places content at the elementary, middle, and high school stages and across all levels, with a column describing how teachers at each level assigned ownership. The third panel, on with whom to share the responsibility, lists parents, counselors, platforms, and districts and policymakers, giving the part teachers assigned to each and the doubt they attached to it.}
\centering
\footnotesize
\setlength{\tabcolsep}{4pt}
\renewcommand{\arraystretch}{1.15}
\begin{tabular}{@{}l p{0.37\textwidth} p{0.42\textwidth}@{}}
\toprule
\multicolumn{3}{@{}l}{\textit{What to teach, and what to leave out}} \\
\midrule
\textbf{Element} & \textbf{What it covers} & \textbf{Where teachers agreed and disagreed} \\
\midrule
Teaching Content & Software that predicts, not a person; who builds it and why; what is safe to share & Named first by nearly every teacher; folded into existing internet safety lessons \\
 & Using without depending; getting help rather than answers; the need behind the turn to the bot & Framed as caution by most, as reflection by some secondary teachers \\
Teaching Format & Discussion of concrete cases, including the scenario cards; inside existing slots; repeated over time & Almost no one wanted a standalone course \\
Teaching Exclusions & \textit{Romantic Partner} for younger grades; \textit{Parent} as too private for a classroom & Withholders cited contagion and stigma; others would withhold nothing \\
\midrule
\multicolumn{3}{@{}l}{\textit{When to teach: across grade levels}} \\
\midrule
\textbf{Stage} & \textbf{Content teachers placed there} & \textbf{Ownership pattern} \\
\midrule
Elementary & The companion is a computer, not real; what is safe to share & Elementary teachers assigned ownership onward to middle or high school \\
Middle & Relationships and warning signs; where \textit{Romantic Partner} content begins & Named by every level as the pivotal stage, yet claimed by few \\
High & Judgment: deciding what role AI will play in one's own life & High school teachers pointed back to earlier stages \\
All levels & A spiral modeled on sex education, repeated at every stage & Framed as a district responsibility requiring vertical alignment \\
\midrule
\multicolumn{3}{@{}l}{\textit{With whom to share: across stakeholders}} \\
\midrule
\textbf{Stakeholder} & \textbf{Part teachers assigned} & \textbf{The doubt attached} \\
\midrule
Parents & Primary responsibility, since use happens at home; should be informed and taught alongside students & Described as unaware, overstretched, or absent \\
Counselors & Default destination for every relational case & Stretched thin; know only what teachers bring them; referral fails without rapport \\
Platforms & Alerts when a boundary is crossed, aggregate rather than per-chat visibility, verified age gates & Little trust in developers whose incentive is engagement \\
Districts and policymakers & A chain of command, grade-level standards, clarity about limits, training & The needed guidance does not exist; decisions left to individual teachers \\
\bottomrule
\end{tabular}
\end{table*}

\subsection{What to Teach in AI Companion Literacy Units, and What to Leave Out}\label{sec:teach}
Teachers were also asked what a short unit on AI companion literacy, taught to every student at their level, should contain. Their answers addressed three questions: the teaching content, where they converged on two lessons regardless of grade level; the teaching format, where they overwhelmingly preferred discussion of concrete cases over direct instruction; and the teaching exclusions, where they divided over whether anything should be deliberately withheld.

\subsubsection{Teaching content: what the companion is, and the difference between use and reliance}
The first lesson nearly every teacher named was that the companion is software. P1, also a media specialist who helps write his district's AI policy, said, \textit{``The very first lesson that I would want to teach is helping the students understand that AI is not necessarily a companion. It can feel like a friend. But ultimately, it's a computer program that looks at patterns and predicts what it should say.''} High school teachers added that this cannot be assumed: \textit{``We think students will think, oh, they'll know that that's not real, but I don't think they do''} (P29). Several wanted students to know not only what the system is but whose it is: \textit{``Who is building this? Who runs Chat, and who runs Claude, and what is their ultimate goal?''} (P29). Elementary teachers folded the topic into internet safety lessons they already teach: \textit{``What's safe to tell AI and what's not safe to tell AI. Just like when you're chatting with people on a website, you don't give them your phone number, you don't give them your address''} (P11). P12 described the progression: \textit{``In the lower grades, your focus is on your sharing of that personal information, and then as you progress through each grade, you're going to bring in that relationship factor.''}

The second lesson was the difference between using a companion and depending on it, and, for the academic scenarios, between getting help and getting answers. P14 would teach \textit{``what a healthy relationship to AI technology looks like, and what a dangerous relationship with AI looks like.''} P3 put it in the terms she uses with elementary students: \textit{``You can't be on it all the time, and it doesn't replace a human. You can't fall in love with it, you can't depend on it as much as you probably want to.''} P31 would teach balance through disagreement: \textit{``If the companion never disagrees with you, then you may have a negative reaction in the future with a human that could not prepare you for that.''} A smaller group of secondary teachers wanted the lesson to be reflective rather than cautionary. \textit{``The bigger question is, what is it that you can tell the bot that you can't tell a friend, or you can't tell a human? What are you afraid of?''} said P29, distinguishing emotional need, which \textit{``needs to be addressed with a human,''} from academic need, \textit{``more of a positive need.''} P33 said that if students \textit{``feel like their friends are gonna judge them, then maybe we need to have a bigger conversation about who they're friends with.''} For \textit{Tutor} and \textit{Study Buddy}, the content was what teachers already teach: \textit{``using it for help, not to do all your work for you''} (P8); \textit{``What's a hallucination? How can you verify if the information they give you is correct?''} (P27); \textit{``Students have to know the skill set before they can go to AI to get help''} (P29).

\subsubsection{Teaching format: concrete cases inside existing slots, not a standalone course}
Teachers overwhelmingly preferred discussion of concrete cases over direct instruction, and many proposed reusing the scenario cards themselves. P28 sketched a lesson: \textit{``Before they can apply something personally, they have to explore it impersonally. I would build a lesson first using objective, impersonal stimuli, like these scenarios, and then I would have them figure out what their intervention would be for Lynn, Sam, or Jordan. That is way more useful than me showing them slides.''} \textit{``I would love to teach directly about all of these scenarios, what are positives, what are negatives, and let's debate it''} (P22). Others reached for real cases: P23 would use \textit{``case studies where AI relationship companions may have been harmful.''} P20 rejected packaged content: \textit{``I don't want videos, I don't want stuff explained by teenagers. People need to have organic conversations with these things.''} Almost no one wanted a standalone course. Teachers placed the unit inside health (P14, P29), English (P15, P32, P33), digital citizenship (P12), a technology special (P9, P19), or counselor-led lessons (P17, P26), and stressed repetition: \textit{``It needs to be reiterated multiple times. As time moves on, you get more news stories''} (P17).

\subsubsection{Teaching exclusions: what should be deliberately withheld}
Exclusions followed the same academic-to-intimate gradient as teachers' concerns. \textit{Romantic Partner} was the most commonly withheld, always on grounds of age: \textit{``No flirting, I would leave that out,''} said P3, assigning it to \textit{``the middle school teacher''}. A few would also withhold \textit{Parent}, because it might \textit{``diminish the relationship with a parent''} (P12) or because home life is too private for a classroom (P15). Two further reasons cut across scenarios. The first was contagion: \textit{``If you're proactive with this, you might give kids a bad idea''} (P19); teaching it \textit{``can oddly backfire and lead a student to be more likely to try that out, because they learn about it in school''} (P22). The second was stigma: \textit{``It's gonna come out that somebody thinks this idea is funny, and that the person that uses these things is lonely and to be made fun of''} (P15). Not all teachers accepted these exclusions. \textit{``I would not leave anything out. I would rather have all the risk put out there. The more information, the better''} (P4). P15, drawing on training as a social-emotional learning coach, rejected the contagion argument directly: \textit{``With sex education, we have the statistics that show that students that are exposed to information that's open with them tend to use the knowledge to create healthy boundaries for themselves.''} P33 agreed: \textit{``If you're leaving anything out, you're giving it an air of mystery, which makes them want to look it up more.''}

\subsection{When to Teach AI Companion Literacy: Sharing Responsibility Across Grade Levels}\label{sec:stages}

\subsubsection{Start early and repeat at every stage}

Teachers agreed that AI companion literacy should begin early and be revisited at every stage, and most described the sequence in the language of sex education or health class, taught age-appropriately from the earliest grades onward (P30). P16 laid out the model: \textit{``Elementary school is when they start talking to you about puberty. Middle school is when you start raising the ante and you become aware of applications and websites. In high school you get a little bit more serious, maybe you investigate real life examples.''} The content at each stage was consistent across teachers: in early grades, that AI is a computer and what not to share (P1: \textit{``by the time they leave elementary, they should know it's not real, it's a computer''}); in middle school, relationships and warning signs, the point at which talk of relationships \textit{``lends itself really quickly to companionship online''} (P16); in high school, judgment (P22: \textit{``now you're gonna decide for yourself. What role is AI gonna play in your life?''}). \textit{Romantic Partner} was reserved for middle school or later by nearly everyone (P24), while \textit{Study Buddy} and \textit{Tutor} were placed wherever devices arrive. Repetition mattered more than placement: \textit{``Teaching takes redundancy. We can't expect them to hear it one time and then go, oh, OK''} (P24); \textit{``the second you stop teaching them and repeating it they're going to just run with it''} (P9).

\subsubsection{No agreement on who owns it}
Agreement ended at ownership. Of 33 teachers, 10 assigned primary responsibility to their own grade level, 12 assigned it to a different one, 8 said every level shares it equally, and one (P25) said it should not be teachers at all; two gave no clear answer. The direction of assignment was consistent: elementary teachers pointed to middle or high school (P4: \textit{``I would put the onus on middle school''}; P8: \textit{``at the start of tomorrow, I'd say high school''}), and high school teachers pointed back (P24: \textit{``elementary school needs to take the lead''}). Middle school was the stage most often named, by teachers at every level, as the point where \textit{``social dynamics get harder''} (P20) and \textit{``kids retreating into the self''} begins (P28). Teachers who claimed the work for their own level gave the same reason: \textit{``The middle school teacher has a very unique responsibility. They're present in the students' lives at a time when they can have the most impact on this''} (P14). Those who said every level shares it framed it as a district responsibility requiring \textit{``vertical alignment''} with a district roadmap (P29), because \textit{``every grade level has the same responsibility, it just looks like different things for all of them''} (P22).

\subsection{With Whom to Share the Responsibility: Parents, Counselors, Platforms, and Districts}\label{sec:stakeholders}

\subsubsection{Parents hold primary responsibility, but teachers doubted they would exercise it}
Parents were named first by most teachers, because companion use happens at home: \textit{``It's primarily the parents, because it is being used at home. They're the ultimate ones''} (P1). Almost in the same breath, teachers described parents as unaware, overstretched, or absent: \textit{``Parents are overworked, stretched to the limit. But ultimately, when you choose to have children, that's your responsibility. Will they do it? A lot of times, no''} (P24); \textit{``I imagine mom has no idea that he has a fake mom. So how many parents of my own students right now are completely unaware of their students' interactions with AI?''} (P28, \textit{Parent}). What teachers wanted, accordingly, was for parents to be informed and equipped, whether through lessons held \textit{``after school with parents, so that both are partaking in the process of becoming more literate with AI''} (P12) or simply because, as P24 put it, parents need to be taught what she was herself still learning.

\subsubsection{Counselors receive the relational cases, and know about them only if teachers tell them}
Counselors were the default destination for every relational scenario, for reasons of training and of role: \textit{``A counselor is kind of built for that social-emotional part for the student body. But again, the counselor won't know until the teacher brings it towards them''} (P17). Teachers also knew the counselor's capacity was limited: \textit{``We've got between 600 and 700 kids and 2 counselors. So they're stretched thin''} (P1), a shortage P33 echoed at a school of 2,500 students. P16 added that referral itself can fail for a child who sees the counselor \textit{``three times every school year,''} because \textit{``it's very easy for a child to lie to an adult that they don't have a close relationship to.''}

\subsubsection{Platforms should make concerning use visible to adults and keep children out of intimate roles}
Teachers asked two things of companies. The first was visibility: \textit{``some sort of protection in the system that would let someone know that some sort of boundary has been passed, that we've reached a moment where a student is in trouble''} (P6). P1 wanted aggregate rather than individual visibility, \textit{``Not that they should see every chat. If you have 20 kids, and 11 of them asked about the theme of Great Gatsby, that's where I think the teacher should be involved on the back side,''} and P21 already used a classroom tool that \textit{``will alert me if they're saying anything concerning.''} The second was age-appropriate design: verified age gates (P2), safeguards keeping minors out of romantic roles (P19), and limits on time as well as age (P31). These requests came with little trust: \textit{``I don't know if I have blanket trust in an AI developer to ethically build out my students' social-emotional skills, when the developer stands to benefit from increased usage of the tool''} (P28).

\subsubsection{Districts and policymakers should supply the protocol and training that teachers lack}
Nearly every teacher said the guidance they would need does not exist: \textit{``There are no policies on these types of things. It shouldn't be up to me, it should really be a district decision. But unfortunately, districts aren't making those decisions''} (P29). What they asked for was concrete: a chain of command (P12), standards by grade level (P11), clarity about \textit{``where does a teacher overstep their boundaries''} (P24), and training as part of their own professional development (P16): \textit{``Teachers need better training on identifying parasocial relationships with their AIs. If we truly are in the classroom for relationships with students, then maybe this one's more priority than preventing AI cheating''} (P28).

\section{Discussion}
\subsection{From AI Literacy to AI Companion Literacy}
Our participants converged on the outline of a literacy that extends the one schools are already building. Across grade levels they wanted students to know what the counterparty is, a program rather than a person, and one they believed was built to please~\cite{wang2026demand,xiao2026crafting}; what is safe to disclose to it; what dependence looks like, echoing trajectories documented among adult and adolescent users~\cite{laestadius2022too,xie2022attachment,namvarpour2026understanding}; and what unmet need lies behind the turn to it, the question P29 called \textit{``the bigger question''} (Section~\ref{sec:teach}). We name this constellation \textit{AI companion literacy}: the capacity to recognize, interpret, and manage relationships with AI systems that occupy human relational roles. It builds directly on the AI literacy that HCI and computing education research have defined~\cite{long2020what,chiu2024artificial,zhang2023integrating,ma2025fostering}, but cannot be simply delivered through the same channel. AI literacy, as currently practiced, is a single-owner curriculum: teachers teach it in their classrooms as part of their subjects. AI companion literacy changes both what is taught, which includes competencies that no single existing instrumental AI literacy fully covers, and who is expected to teach it, which is a negotiation across teachers, counselors, parents, platforms, and policymakers that our participants could outline but not resolve (Section 7). It extends existing AI literacy in three directions.

First, it has an object that no existing literacy covers: a counterparty that is relational but not a person. AI literacy corrects mental models of a tool~\cite{long2020what,touretzky2019envisioning}; social-emotional learning teaches students to sustain relationships with people, whose feelings and stakes are real~\cite{shi2024effective}; digital and media literacies teach them to evaluate content and the human strangers behind it~\cite{difranzo2019social}. A companion falls between these objects: it initiates contact, remembers the student, adapts to them, and, in many commercial systems, is engineered to be agreeable~\cite{muldoon2025cruel,wang2026demand}, the property teachers identified as its central risk (Section~\ref{sec:risks}), though some tutors are explicitly built to challenge rather than agree~\cite{cnn2023khanmigo}. Knowing that the AI is a program was, accordingly, only the first lesson teachers named, because an accurate mental model does not end the relationship: P24 caught herself thanking ChatGPT while knowing exactly what it was (Section~\ref{sec:risks}). The competency teachers' answers point to is managing an attachment that survives the knowledge that its object is software. None of the three parent literacies teaches this: AI literacy stops at the accurate model, social-emotional learning assumes the relationship is with a person, and media literacy assumes the problem is deception and misinformation. Combining them still leaves this gap, which is why we treat AI companion literacy as a new extension rather than a repackaging.

Second, it carries the literacy into a new jurisdiction. AI literacy settled into schools because it could be defined as an academic problem: the New York City reversal resolved student AI use into tool use, integrity, and evaluation, squarely inside teachers' core jurisdiction~\cite{banks2023,whalen2025}, and our teachers reproduced that settlement individually, absorbing \textit{Tutor} and \textit{Study Buddy} with the repertoire they apply to any classroom technology (Section~\ref{sec:how}). The relational content extends into territory teachers share with counselors and parents~\cite{kourkoutas2018teachers,ray2007two}: the content teachers rated most urgent was precisely the content they felt least entitled to teach, and a literacy whose academic portion belongs to everyone and whose relational portion belongs to no one will, by default, be taught only in part.

Third, it inherits the politics of the relational subjects it borrows from. Teachers attached the new content to settlements that exist: health, social-emotional learning, digital citizenship, and above all sex education's age-graded spiral, with the consent and opt-out structures that subject developed over decades~\cite{schmidt2015evidence}. They inherited its central dispute as well: those who would withhold the intimate scenarios reasoned from contagion, those who would withhold nothing cited the sex education evidence that information builds boundaries rather than behavior~\cite{schmidt2015evidence}, and whether comprehensive or restrictive approaches to relational AI content produce better outcomes is an empirical question no one has yet studied.

Naming the literacy matters because the alternative is drift, not because companions are desirable. Companion platforms are already shaping what children learn about relationships, by design and without curriculum~\cite{muldoon2025cruel,yu2026principles}, while the adults around them, parents assessing single events, experts assessing patterns~\cite{yu2026principles}, and our teachers assessing jurisdiction, each see only part of the problem. A defined AI companion literacy gives HCI and education researchers an object to design for and evaluate, gives the frameworks now being written a relational dimension to incorporate~\cite{chiu2024artificial,ma2025fostering}, and gives the adults who share responsibility for it a common term (Section~\ref{sec:rq3}).

\subsection{The Contested Jurisdiction of Teaching in the Age of AI Companions}\label{sec:jurisdiction}

Our findings are difficult to explain as risk assessment and easy to explain as jurisdiction: the scenario teachers judged most harmful drew the least unconditional involvement, and the least harmful drew the most (Section~\ref{sec:rq2}). In Abbott's account, an occupation holds work by diagnosing which problems it is socially entitled to claim~\cite{abbott1988system}, and the three tests teachers applied, visibility, the academic-relational divide, and safety, are diagnostic in exactly this sense: they classify a case as belonging, or not belonging, to the teacher's work. This logic differs from that of the other adults HCI research has studied so far: parents assess single events and experts assess patterns over time~\cite{yu2026principles}, while teachers assess jurisdiction, with harm entering only through the safety test, where the mandated reporter role gives them standing their teaching role does not~\cite{worley2013mandated}. Hence the otherwise puzzling career of the \textit{Parent} card, refused as curriculum yet the clearest case for action of all five: reframed as possible neglect, it left contested relational education for the settled, codified territory of child protection~\cite{walton2022loco}.

For the intimate companions, the result is a jurisdictional vacancy. Teachers assigned primary responsibility to parents they described as unaware, referred cases to counselors they knew were stretched thin, asked for safeguards from companies they did not trust, and waited for district policies that do not yet exist (Section~\ref{sec:rq3}); every adult they named holds a plausible partial claim, and none an established one~\cite{namvarpour2026responsible}. This educational work is declined rather than claimed, passed between occupations and grade levels in a direction that consistently points away from the speaker: teachers at every level agreed the literacy must start early while a majority located ownership elsewhere (Section~\ref{sec:stages}). One reason may be that teachers' authority over students' private lives is negotiated rather than positional~\cite{metz1978classrooms,pace2007understanding}, and claiming this work invites the parental conflict our participants feared. A problem everyone ranks urgent and no one claims is a gap in the division of labor, not in awareness, and awareness campaigns will not close it.

These dynamics carry a practical warning. Without an established claim, the adults with the most consistent access to children default to the one role that is codified: surveillance and referral. AI companion literacy would then be practiced not as education but as screening, a watch for warning signs followed by a handoff, with instruction confined to the academic scenarios that need it least. In practice, this means the students most in need of guidance — those turning to an AI for the emotional support no adult around them provides — receive the least education about what that relationship is. Teachers in our study did recognize this irony: they described the intimate scenarios as the ones students most needed help navigating (Section 5.1.2), yet placed those same scenarios outside their jurisdiction (Section 6.1.2). The concern is genuine, but current occupational arrangements prevent it from crystallizing into instruction. The implications for HCI in Section~\ref{sec:implications} are therefore aimed less at persuading teachers to care (they already do) than at giving their care a jurisdiction to operate in.

\subsection{Implications for Design and Policy}
\label{sec:implications}

We pair each implication below with the tension it raises: several of these directions, monitoring above all, are contested in education, so we flag the questions our data can pose but not settle. Table~\ref{tab:implications} summarizes.

\begin{table*}[t]
\caption{Implications for design and policy (\textbf{Section~\ref{sec:implications}}): each suggestion paired with the tension it raises and the questions our data leave open.}
\label{tab:implications}
\Description{A table with three columns: audience, what we suggest, and the tension it raises together with the questions that remain open. Two rows address platform designers, covering escalation channels rather than open observation and age-sensitive agreeableness; one row addresses schools and districts, covering a codified escalation ladder and a curricular home for AI companion literacy; one row addresses policymakers, covering regulation of the jurisdictional vacancy. Each row pairs its suggestion with a countervailing tension and the specific questions the study cannot settle.}
\centering
\footnotesize
\setlength{\tabcolsep}{4pt}
\renewcommand{\arraystretch}{1.15}
\begin{tabular}{@{}l p{0.30\textwidth} p{0.46\textwidth}@{}}
\toprule
\textbf{Audience} & \textbf{What we suggest} & \textbf{The tension, and what remains open} \\
\midrule
Platform designers & Escalation channels rather than open observation: surface safety-critical moments to a designated adult, not full conversations & Visibility can read as surveillance and chill disclosure. Open: who the designated adult is; what false-alert rate is tolerable; how escalation changes what students say \\
 & Agreeableness as an age-sensitive parameter, with more built-in challenge for younger users & Friction adults endorse may drive students to unmonitored alternatives. Open: how much challenge before abandonment; who sets the dial \\
\midrule
Schools and districts & Codify the escalation ladder teachers improvise; name a curricular home for AI companion literacy & Codified thresholds risk becoming a compliance checklist and one more mandate. Open: the right thresholds; how to formalize without removing discretion \\
\midrule
Policymakers & Regulate the jurisdictional vacancy: not only what companions may say to minors, but which adult is responsible for knowing & Responsibility without resources reproduces the problem as an unfunded screening mandate. Open: which adult the duty should name; what resources must accompany it; how a duty to know is enforced without mandating surveillance  \\
\bottomrule
\end{tabular}
\end{table*}

\subsubsection{For companion platform designers: escalation channels and age-sensitive agreeableness}
We suggest that companion platforms build escalation channels rather than open observation: surfacing safety-critical moments to a designated adult, instead of exposing full conversations. This follows from teachers' account of the central risk, that companion relationships stay invisible until they spill over (Section~\ref{sec:criteria}). The tension is that the same teachers who wanted visibility also rejected an always-on presence as an overreach, and monitoring that students know about may chill exactly the disclosures that make a companion valuable to an isolated student. Open questions include who the designated adult should be, what false-alert rate is tolerable when an alert can trigger a wellness check, and how escalation changes what students are willing to say. We further suggest treating agreeableness as an age-sensitive parameter, with more built-in challenge for younger users, since teachers located much of the risk in chatbots they saw as built to please, and some would teach the value of disagreement (Section~\ref{sec:teach}). The tension is that friction adults endorse may simply drive students to unmonitored alternatives, which leaves open how much challenge a young user tolerates before abandoning a sanctioned tool, and who sets that dial, whether designer, district, parent, or student.

\subsubsection{For schools and districts: codify what teachers improvise}
We suggest that districts codify the escalation ladder teachers already improvise (Section~\ref{sec:how}), with shared thresholds for when noticing becomes a conversation and a conversation becomes a referral, and that they name a curricular home for AI companion literacy (Section~\ref{sec:teach}); only 13 of 33 teachers reported any relevant training, and none reported guidance specific to student-AI relationships. The tension is that codifying relational judgment risks flattening it into a compliance checklist, adding one more mandate for teachers who already described themselves as stretched. What the right thresholds are, and how to formalize response without stripping out the discretion teachers relied on, remain open.

\subsubsection{For policymakers: assign responsibility before harms entrench}
Teachers framed the stakes in generational terms: institutions did not regulate or teach about social media until adolescents had adopted it at scale and its harms had entrenched, and AI companions are now at the equivalent early moment. P1 put it directly: \textit{``I think we dropped the ball on social media, and that's why it's gotten so out of control, but we've got a second chance with AI not to drop the ball this time on it.''} We suggest that policy treat the jurisdictional vacancy our participants described (Section~\ref{sec:jurisdiction}) as an object of regulation: not only what companions may say to minors, but which adult is responsible for knowing. The tension is that a duty assigned without resources becomes an unfunded screening mandate for teachers who lack training, time, and authority, which leaves open which adult the duty should name, what staffing and training would have to accompany it, and how a duty to know can be enforced without mandating surveillance of students' private conversations.

\section{Limitations and Future Work} \label{sec:limitations}
Our findings should be read against several limitations. We interviewed 33 US teachers recruited through Prolific, a sample that skews toward teachers already comfortable with online platforms and cannot represent the full range of US schools, let alone other countries' educational systems. Teachers responded to hypothetical scenario cards, and what they said they would do may differ from what they would do when a real student, parent, and principal are attached to the case. The cards also bundle factors that cannot be separated in our design: the intimate scenarios differ from the academic ones not only in relational role but in setting, legitimacy, privacy, and age sensitivity, so some of the differences we observe may reflect these co-varying features. Disentangling them would require designs that vary such factors independently, for example factorial vignettes. Our account is also one-sided by design: we did not interview the counselors at the end of nearly every referral chain, the parents teachers named as primary, or students themselves, whose views of teacher involvement may diverge sharply from their teachers'. Finally, both companion platforms and the policy landscape are changing quickly, and teachers' judgments may shift as incidents, regulation, and their own AI use accumulate. Future work should follow the referral: studying counselors' capacity and perspective, observing how schools actually handle companion-related cases, and evaluating AI companion literacy curricula.

\section{Conclusion}
Students are forming relationships with AI that take on the roles of tutor, study buddy, friend, romantic partner, and parent. Interviewing 33 US K-12 teachers with scenario cards grounded in real incidents, we found that teachers welcomed the academic companions, worried that the intimate ones remove the friction through which children learn human relationships, and decided whether to act less by harm than by jurisdiction, envisioning the resulting AI companion literacy as work shared with counselors, parents, platforms, and policymakers across grade levels. More broadly, our findings point to a shift in how AI enters education: from a tool students use to a relationship students sustain. This shift changes not only what students need to learn but who is responsible for teaching it, and it asks HCI researchers to design not just for the student-AI interaction but for the institutional infrastructure around it. Building this literacy will require more than adding content to AI literacy frameworks: it will require deciding whose work students' relationships with AI are. Teachers in our study were willing to notice, to teach, and to refer. What they lacked was not care but a jurisdiction in which their care could operate.

\bibliographystyle{ACM-Reference-Format}
\bibliography{refs}


\end{document}